# Extensive air showers and atmospheric electric fields. Synergy of Space and atmospheric particle accelerators.

A. Chilingarian

A. Alikhanyan National Laboratory (Yerevan Physics Institute), Alikhanyan Brothers 2, Yerevan, Armenia, 0036

**Abstract**

Various particle accelerators operate in the space plasmas, filling the Galaxy with high-energy particles (primary cosmic rays). Reaching the earth's atmosphere, these particles originate extensive air showers (EASs) consisting of millions of elementary particles (secondary cosmic rays), covering several km$^2$ on the ground. During thunderstorms, strong electric fields modulate the energy spectra of secondary particles and, consequently, originate short and long particle bursts. Impulse amplifications of particle fluxes (the so-called thunderstorm ground enhancements, TGEs) manifest themselves as peaks in the time series of count rates of particle detectors, which coincide with thunderstorms, during which free electrons are accelerating and multiplied, forming electron–gamma ray avalanches. Thus, electron accelerators emerging in thunderous atmospheres can significantly alter the frequency of EAS triggers. Free electrons, abundant at any altitude in the atmosphere from the small to large EASs, serve as seeds for these accelerators. EAS cores randomly hitting arrays of particle detectors also generate short bursts of relativistic particles. The lightning flashes abundant on mountain altitudes also possibly generate additional seed electrons. To understand the complicated interplay of secondary particle fluxes, electric fields, and lightning flashes, particle detectors, and spectrometers have operated for many years on mountain peaks, providing valuable information for establishing a field of high-energy physics in the atmosphere.

**Plain language summary**
Correlated measurements of particle fluxes modulated by strong atmospheric electric fields, registration of broadband radio emission from atmospheric discharges, and registration of various meteorological parameters lead to a better understanding of the complex processes of particle-field interactions in the terrestrial atmosphere. The cooperation of cosmic rays and atmospheric physics has led to the development of models for the origin of particle bursts recorded on the earth's surface, vertical and horizontal profiles of electric fields in the lower atmosphere, the modulation of particle fluxes in the atmospheric electric fields, etc. Interdisciplinary atmospheric science primarily requires monitoring particle fluxes around the clock by synchronized networks of identical sensors that record and store multidimensional data in databases with open, fast, and reliable access. The advances in multidimensional measurements over the past decade significantly intensified the development of new models of atmospheric electricity, giving more insight into understanding EAS development in the strong atmospheric electric fields.

## 1. Introduction

Millions and millions of particles are sent toward the Earth's surface by the most powerful natural particle accelerators operating in the Universe and the electrified atmosphere. Protons and fully-stripped nuclei, which are accelerated in violent explosions in our galaxy and beyond, enter the Earth's atmosphere and unleash gigantic particle cascades. Particle detectors arranged in large surface arrays observe particle bursts, which are analyzed to determine the sources of primary nuclei and the acceleration mechanism. Modern large EAS experiments are located in high-altitude regions with frequent thunderstorms, which induce strong electric fields aloft the detector. Atmospheric electric fields have a significant impact on the development and structure of particle showers. They affect the shower in several ways, including the acceleration and deceleration of electrons and muons; the Lorentz forces bend the trajectories of charged particles, influencing the lateral and longitudinal spread of the air shower. The particle density within the shower is also affected by the charge separation within the EAS. Furthermore, strong atmospheric electric fields can lead to the emergence of relativistic runaway electron avalanches (RREAs [1-4]), causing a sizeable impulsive enhancement of the number of electrons and gamma rays in the EAS.

Understanding the relationship between cosmic rays and atmospheric electric fields is crucial for accurately identifying the type and energy of primary particles. Consequently, scientists conduct intensive simulations to study any biases introduced by atmospheric fields in the basic physical inferences obtained from EAS experiments. Studying the interactions between cosmic rays and the Earth's atmosphere provides valuable insights into fundamental atmospheric processes. A comprehensive approach, including particle and atmospheric physics instrumentation, can reveal the reasons behind a significant increase in EAS trigger frequency during thunderstorms. This approach can also help to understand the complex atmospheric processes that cause these modulation effects and generate large particle bursts. This review investigates the relationship between extensive air showers (EASs), lightning flashes, and the intriguing processes of high-energy physics in the atmosphere (HEPA, [5]). We intend to shed light on the complex interplay between these phenomena and provide insights into the underlying mechanisms of particle burst phenomena.

## 2. Change in the frequency of EAS triggers during thunderstorms.

Figure 1 illustrates the particle fluxes that reach the Earth's surface initiated by the space and atmospheric accelerators and Radon progeny gamma radiation. The left side of the diagram shows the electron avalanches that develop in the lower and upper dipoles of the thundercloud, TGEs, and TGFs, respectively. Lightning flashes reduce the negative charge above the Earth's surface, decreasing the electric field in the lower dipole below the RREA initiation threshold. This decrease leads to a decline in RREA, eliminating high-energy particles from the TGE flux. However, TGE continues even after the near-surface electric field strength returns to the fair-weather value due to the tens of minutes-long lifetime of non-stable isotopes 214 Pb and 214Bi. On the right side of the cartoon, we show the gamma-ray and proton originated in the supernovae remnant and initiated EASs in the terrestrial atmosphere.

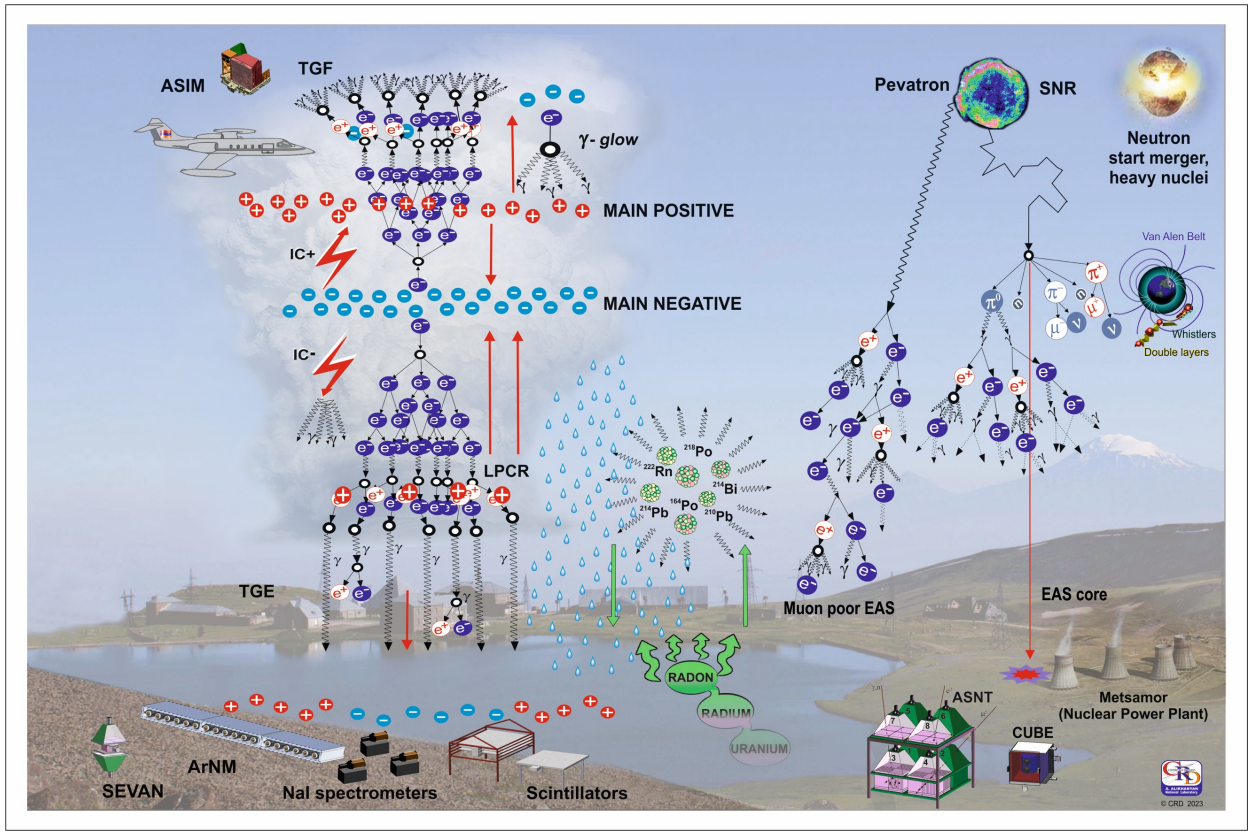

**Figure 1. The fluxes of secondary particles from space and atmospheric accelerators, as well as gamma radiation from 222Rn progeny. The cartoon also shows the charge structure of a thundercloud, the direction of electric fields, lightning flashes, and various measuring facilities on the surface and in space. Additionally, we can see the sources of primary cosmic rays. In the background is the Aragats cosmic ray station, located at 3200 meters and equipped with various particle detectors and spectrometers.**

Suppose we merge all three sources of secondary particles, as shown separately in the cartoon for clarity. In that case, we will see that the electrons from the EAS will be accelerated and multiplied in the atmospheric electric field. This will increase the count rates of particle detectors and the surface array trigger frequency, affecting the energy estimates obtained from the shower size.

Early measurements of the EAS-TOP detector in Italy [6] revealed a significant excess of the EAS triggers lasting 10–20 minutes during thunderstorms. Recently, high-altitude experiments have paid attention to possible biases due to the propagation of EASs through strong atmospheric electric fields. The ARGO-YBJ experiment [7] is well suited for researching the modulation effects electric fields pose on EAS due to its high-altitude location at the Tibet plateau (4300 m), where thunderstorms are frequent. Also, the central full-coverage carpet allows particle density measurement with high precision. At ARGO-YBJ, two data acquisition systems (the scaler and shower operation modes) worked independently. In scaler mode, the counting rate of each detector cluster was measured every 0.5 s. In shower mode, the detector was triggered when at least 20 pads in the central carpet were fired within 420 ns. A 20% enhancement of the EAS triggers was detected in 2012. The mean duration of 20 episodes of enhanced trigger rates coinciding with thunderstorms was ≈30 minutes. The Monte Carlo simulation, performed for different strengths and polarities of the field (up to about 0.2 kV/cm), satisfactorily describes the observations regarding the amplitude and sign of the trigger rate variations. Simulations were performed assuming a uniform electric field 500 m above the detector. Trigger rate variations depend on the near-surface electric field (NSEF) sign, measured by the BOLTEK's electric mill EFM 100. If NSEF is positive, the count rate decreases; if negative – it increases (see Fig. 1, 2 of [2]). The trigger rate was smoothly changed with changing NSEF during 1.5 hours of the storm. They also noticed that the NSEF has more effects on EAS triggers corresponding to larger zenith angles.

The LHAASO [8] surface array is located at Haizi Mountain, Daocheng County, Sichuan Province, which is at the edge of the Tibetan Plateau with an altitude of up to 4410 m. The Tibetan plateau is known for frequent thunderstorms and large intracloud electric fields, whose vertical profile can extend to 1-2 km. The largest component of LHAASO, KM2A, contains 5216 electromagnetic particle detectors (EDs) and 1188 muon detectors (MDs). The ED array covers an area of 1.3 $km^2$ in a triangular grid. The EDs detect the electromagnetic particles in the shower, which are used to reconstruct the shower core position, arrival direction, and primary particle energy. EAS trigger logic requires at least 20 EDs fired within a time window of 400 ns. The DAQ records 10 μs of data from all EDs and MDs for each shower event with signals over the thresholds. From October 2019 to March 2022, more than 200 thunderstorm events were detected by the BOLTEK's field mill EFM 100 at the LHAASO observatory. Most of them were recorded from April to September. A significant increase was observed in the number of particles (EAS size $N_e$) detected by the ED array (≈20%) and in the trigger rates (≈20%).

The High Altitude Water Cherenkov (HAWC) observatory is in the saddle region between the Sierra Negra and Pico de Orizaba in Mexico. It is located at 4100 meters above sea level. Due to its altitude, the observatory is affected by electrically charged clouds and lightning flashes for approximately six months of the year. HAWC consists of 300 water Cherenkov detectors (WCD), instrumented with four photomultiplier tubes (PMTs) and spread over 20,000 $m^2$. Each WCD is 7.3 m in diameter and 5.0 m in depth. The main DAQ measures arrival times and time over thresholds of PMT pulses. It also has a TDC scaler system that counts the hits within a time window of 30 ns of each PMT and the coincidences (multiplicities) of 2, 3, and 4 in each WDC. Recently, a small, fast scintillator detector (7.62 × 7.62 cm LaBr3) was located near giant water-Cherenkov detectors of HAWC. The detector output was attached to the Broadband Interferometric Mapping and Polarization (BIMAP) sensor's electronics. BIMAP captures 15 ms of data for each trigger with 5 ms of pre-trigger data. This detector observed particle bursts between September 2017 and September 2019 during fair-weather days, meaning there were no nearby lightning flashes, see Tab. 1 of [10]. CORSIKA [11] simulations confirm that particle bursts originated from EAS core particles captured in nuclei of soil, which produced high-energy gamma rays through (n, γ) reactions.  HAWC's particle bursts were not related to atmospheric discharges; the observed bursts are initiated by particles from EAS cores hitting the HAWC array.

Covering an area of 3000 km$^2$, the Pierre Auger Observatory represents the largest ground-based experiment for identifying ultra-high energy cosmic rays located at Argentinian Pampa. The surface detector, which consists of 1660 water-Cherenkov detectors, samples the footprint of the air showers as they hit the ground, while the FD, with its 27 telescopes distributed over four sites on the perimeter of the SD, records the longitudinal development of EAS n the atmosphere by registering the fluorescent light. During thunderstorms, twice a year, the SD observed peculiar events correlated with thunderstorms that were different from an extensive air shower. A typical footprint of these events covers about 200 km$^2$ (much larger than EAS events) and lasts more than 20 μs, an order of magnitude longer than EAS events. However, a low detection rate may indicate that events were not caused by physical origin but rather by the trigger optimized for EAS events.

Another large detector, the surface detector of the Telescope Array (TA) experiment (TASD, [14]), comprises 507 scintillator detectors on a 1.2-km square grid occupying a total 700 km$^2$ area. TASD provides shower footprint information, including core location, lateral density profile, and timing, which are used for recovering shower axes and energy. Each measuring unit comprises upper and lower scintillators 1 cm thick and a 3 m$^2$ area. A 1-mm-thick steel plate separates the upper and lower planes. They are read out by photomultiplier tubes coupled to the scintillator via an array of wavelength-shifting fibers. A 12-bit ADC digitizes the output signals from the photomultipliers with a 50-MHz sampling rate. An event trigger (frequency 0.01 Hz) is recorded when three adjacent units observe a signal larger than three vertical equivalent muons (VEMs) within eight μs. (≈ 2 MeV per 1 cm of scintillator). From the changing count rates (± 1-2%) of the TA scintillators, one can see a clear correlation between a deficit or an excess in the CR intensity with a thundercloud movement [15]. The intensity variations "move" in the same direction as the thundercloud for tens of minutes at a ∼ 2 km/min speed. These variations correlate with lightning flashes and atmospheric electric fields without lightning. The electric field magnitude found to reproduce the observed intensity variations was approximately 0.2-0.4 GV.

Consecutive TASD triggers were recorded in 1-ms intervals correlated with lightning flashes above the TA detector, observed by the Lightning Mapping Array (LMA) and the Vaisala National Lightning Detector Network (NLDN). The Lightning flashes that produced trigger bursts were very rare. There are typically about 750 NLDN-recorded flashes (IC and -CG) per year over the 700-km$^2$ TASD array. In 8 years of TA operation, were detected only 20 bursts correlated with lightning activity. Thus, fewer than 0.5% of NLDN flashes recorded over the TASD were accompanied by identifiable gamma bursts. The burst durations were within several hundred microseconds, and the altitude of the source was typical of a few km above ground level. In [16], the authors introduce a lightning-related scenario of the burst origination: "The results show that the TGFs occur during strong initial breakdown pulses (IBPs) in the first few milliseconds of -CG and low-altitude intracloud flashes and that the IBPs are produced by a newly-identified streamer-based discharge process called fast negative breakdown."

In the late 20th century, the MAKET ANI surface array (3200 m, Aragats mountain in Armenia [17]) became the first instrument to measure the energy spectra of both light and heavy galactic nuclei separately. The results showed a very sharp knee in the light component at 2-3 PeV, while no knee was observed in the heavy component up to 10 PeV. We kept running sixteen 1 m$^2$ area and 5 cm thick plastic scintillators of the array and registered two large TGEs on 19 September 2009 [18] and 4 October 2010.[19] The trigger condition was the detection of signals minimum in 8 scintillators within 1 μs. At the minute of maximum TGE flux, the enhancement of the MAKET triggers was 250% (see Fig. 2 of [19]). MAKET array produces two sets of samples composed of strings of numbers that represent the registered particles in each of the 16 scintillators. One set contains pure background triggers from EASs, while the other includes triggers from RREAs contaminated by background (EASs). Our challenge is to purify the signal sample by enriching it with RREA events. I.e., selecting the decision boundary in the parameters space to select maximum RREA events and suppress TGE events as much as possible. We used the mean and maximum particle densities in 16 scintillators. The densities of TGE particles originating from multiple RREAs in the thundercloud 1-2 km above with maximal energy not exceeding 50 MeV are expected to be uniform without any significant peaks. The typical particle density distribution of the EAS hitting Earth's surface is bell-like, with a substantial fraction of the shower particles near the EAS core. Thus, if the EAS core is near scintillators, enormous densities can occur, and the mean density is expected to be larger than one from RREA. The linear discrimination function in 2-dimensional feature space suppressed ≈50% of the "pure" EAS events, losing ≈25% of the joint EAS and TGE events (148 from 613, see Figs. 3, 4 of [19]). The 25% contamination could not be significantly reduced due to large EASs with axes far from the MAKET array. The long tails of EASs generate events with low mean and maximal densities that could not be distinguished from densities generated by RREAs. After analyzing 465 selected ECSe (Extensive Cloud Showers), which are also called microbursts by Alex Gurevich [20], we found out that the ECS particle density follows an exponential law and that the ECSs are distributed within a 100-meter radius, limited by the size of MAKET array.

3. **Particle bursts and lightning activity**

Thunderstorms create strong electric fields within and around the storm system. These electric fields result from charge separation within the storm clouds and between the clouds and the ground. In-cloud and cloud-to-ground lightning lower the potential difference between charged layers of the thundercloud and stop electron acceleration. In 1945-1949, Joachim Kuettner conducted groundbreaking experiments at Zugspitze [21], where he discovered the tripole structure of the atmospheric electric field. This model suggests that the atmospheric electric field comprises symmetric upper and lower dipoles, accelerating free electrons toward the open space and the Earth's surface. Electrons accelerate into space and create avalanches in the upper atmosphere, resulting in bremsstrahlung gamma rays. The most energetic gamma rays occasionally reach orbiting gamma observatories, creating short bursts of particles known as terrestrial gamma flashes (TGFs [22,23]). The accelerated electrons in the lower dipole create particle avalanches that register on the ground as thunderstorm ground enhancements (TGEs [18,19]). These TGEs consist of millions of gamma rays, fewer electrons, and rarely neutrons. The RREA process results in the production of both TGFs and TGEs. This process begins when the strength of electric fields exceeds a specific critical value for the air density. Balloon experiments conducted in New Mexico [24] and TGEs detected on Aragats [25,26], Musala, Lomnicky Stit mountains [27], and other places have proven this. Numerous simulations have also demonstrated the exponential increase in particle numbers after the modeled electric field surpasses the critical value at distances of 1-2 km. There is a comprehensive compilation of information on TGE physics in the recently published catalog [28].

Several research papers [13,15] introduced a different name for the particle bursts observed on the Earth's surface - Downward Terrestrial Gamma Flashes (DTGFs). These papers claim that the RREA process, responsible for TGEs, also plays a role in the origination of DTGFs. However, the seed electrons for DTGFs are not supplied from EASs but from atmospheric discharges or through a feedback process. Both provide RREA with additional free electrons, producing extremely bright particle bursts. Therefore, the critical difference between TGFs and TGEs lies in the different populations of seed electrons. Thus, the crucial for proving the DTGF origination is detecting the lightning flash at least "several hundred μs" before detecting the particle burst. However, recent measurements of the Atmosphere-Space Interactions Monitor (ASIM) on board the International Space Station contradict the "lightning" scenario of TGF origination. Measurements of TGFs, taken in conjunction with the optical detection of lightning flashes by the ASIM instruments, suggest that TGFs occur before the onset of optical pulses [29]. TGFs do not co-occur after or simultaneously with lightning flashes but approximately 1.4 ms before. In [30], an analogical scenario for TGF production was introduced. Namely, the gamma rays are produced several ms before a narrow bipolar event (NBE), which often marks the lightning initiation. Furthermore, analysis of four TGF catalogs from different instruments revealed that a significant proportion of TGFs lead to increased lightning activity detected in radio waves (spherics) between 150-750 ms after TGFs occur [31]. This suggests that TGFs can act as precursors of lightning and initiate lightning flashes, whereas lightning activity does not produce any additional seeds for Relativistic Runaway Electron Avalanches (RREAs).

Simply put, the statement that TGF occurred after the return stroke began is at odds with ASIM measurements [29]. Also, the analysis of TGF catalogs from major gamma-ray orbiting laboratories indicates that increased lightning activity detected in radio waves (spherics) occurs after TGFs have occurred [31] and not before. Another scenario of DTGF origination, the feedback process in which positrons and gamma rays from RREA avalanche travel back up to the thundercloud and restart RREA, was never measured in the experiment and has not been confirmed in the TGE detection [32].

## 4. TGEs and TGFs

The 24/7 monitoring of particles, fluxes, atmospheric discharge, and near-surface electric fields on Aragats provides compelling evidence that TGEs are precursors to lightning flashes. During the first TGE research campaign on Eastern Europe and German mountaintops, it was observed that lightning flashes stop the particle fluxes and never initiate them. During episodes of enhanced particle fluxes lasting several minutes, lightning activity is suppressed. Therefore, TGEs and TGFs appear to be the same phenomenon developing within large-scale strong atmospheric electric fields symmetrically aligned within the lower and upper dipole of a tripole structure of the electric field. Lightning initiation involves RREAs, which ionize the air and create a channel for the lightning leader to follow. TGEs observed on the Earth's surface and TGFs observed by the orbiting gamma observatories are manifestations of the sizeable ionizing electron flux in the thundercloud produced by the RREA electrons.

The difference between TGFs, which last for a few milliseconds, and TGEs, which last for minutes, is due to the distance between the particle source and the detector. TGEs are detected ≈1000 times closer to the particle source than TGFs. Only the most energetic gamma rays from RREAs can reach the Space Station 400 km from the Earth, producing a short particle burst, sometimes followed by a few more rounds (multi-pulse TGFs registered by ASIM, see figures in [33]). On Aragats, RREAs create an almost continuous flow of particles because of the proximity to the source detectors exposed to the particle beams from the in-cloud accelerator.

In Fig 2, we compare the 10 ms duration time series of the multi-pulse TGF [33], registered by ASIM, the detector specially designed for TGF detection, and the same duration time series from the STAND1 detector, which registered a large TGE on May 23, 2023. ASIM detector is much more efficient in registering gamma-ray bursts than other orbiting gamma laboratories designed to detect gamma rays from violent explosions in the Universe and use complicated off-line triggers to find TGFs from the Earth's direction.

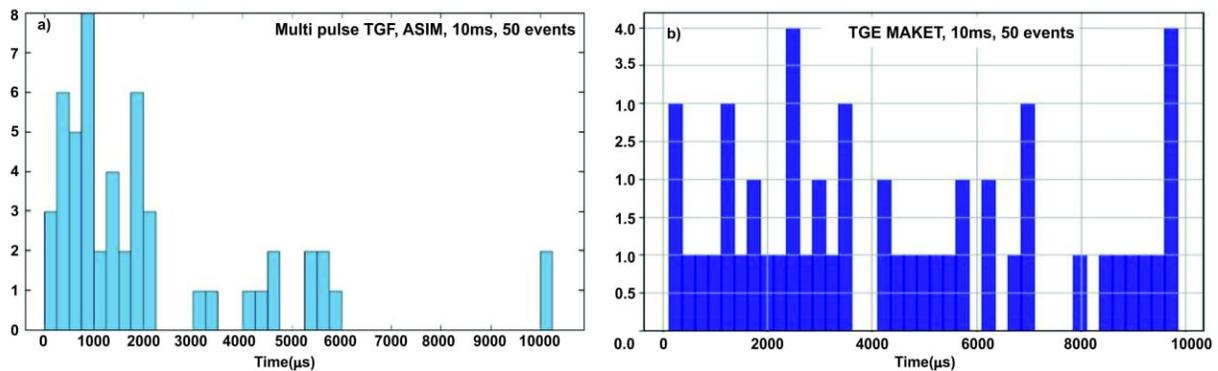

**Figure 2. a) Time series of gamma-ray arrival times from the onset of the first TGF until the onset of the last TGF, registered by ASIM's HED detector; b) Time series of the TGE particle arrival times registered by STAND1 (MAKET); zero time corresponds to 00:34:53.1 on May 23, 2023. The bin width of both is 250 μs.**

The light pulse from ASIM's BGO is several hundreds of ns long; therefore, it has an effective time resolution of about 1 μs and a size of 900 cm$^2$. The STAND1 detector attached to a digitizing oscilloscope registers the time series of the TGE particle arrival times. It is 3 cm thick, 1 m$^2$ area upper plastic scintillator of the stacked detector array. The full width on half maximum of the photomultipliers pulse (FWHM) is 30-40 ns, the sampling rate of the digitizing oscilloscope is 4 ns, and the detector size is 10,000 cm$^2$. Both detectors register 50 particles in 10 ms. However, the particle flux density at ASIM is ≈ ten times larger due to its smaller size. Fig 2a shows a very intense gamma-ray burst (left side) followed by a few discrepant gamma rays. In Fig 2b distribution of particles is more-or-less uniform.
Figure 3a shows the extended to 200 ms time series of the STAND1's TGE detection. Figure 3b shows a time series obtained with randomly generated 992 events (the same number of events as in 3a) from the Poisson distribution. The simulation aims to demonstrate that TGE particles come randomly from multiple avalanches in a large-scale thunderous atmosphere. Instead, TGFs comprise a few most energetic avalanches, occasionally reaching the Space Station at a distance of ≈ 400 km from the source.

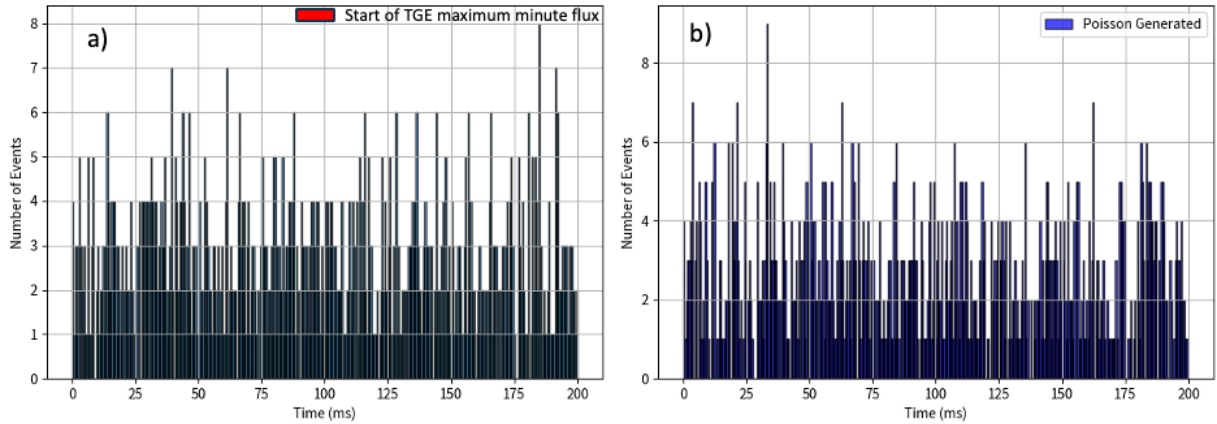

**Figure 3. a) 200 ms time series of the TGE particle arrival times registered by STAND1 (MAKET), zero time corresponds to 00:34:52.9 on May 23, 2023. b) time series of the Poisson random times within 200 ms. The bin width of both is 500 μs.**

The maximum period of multi-pulse TGF shown in [33] is 10 ms. However, the balloons or aircraft flying above thunderstorms in equatorial regions will observe minutes-long TGFs comprising millions of particles. The 23 May TGE continued another minute till a lightning flash abruptly terminated it. Figure 4 shows the 50-ms time series of count rates of 1 cm thick and 1 m$^2$ area plastic scintillator from STAND1 array near GAMMA experimental hall on Aragats. The enhancement of the scintillator count rate was ≈10-fold, increasing from 22 to 240 and after a nearby lightning flash in 100 ms, decreasing to the previous level (see details in [34]). During TGE's highest flux, lightning activity was suppressed, and TGE reached the maximum flux intensity. As shown in Figure 4, there were three attempts to start TGE. The first two were interrupted by a lightning flash in the initial stage of TGE, and only the third reached a record count rate. WWLLN detected all three lightning flashes.

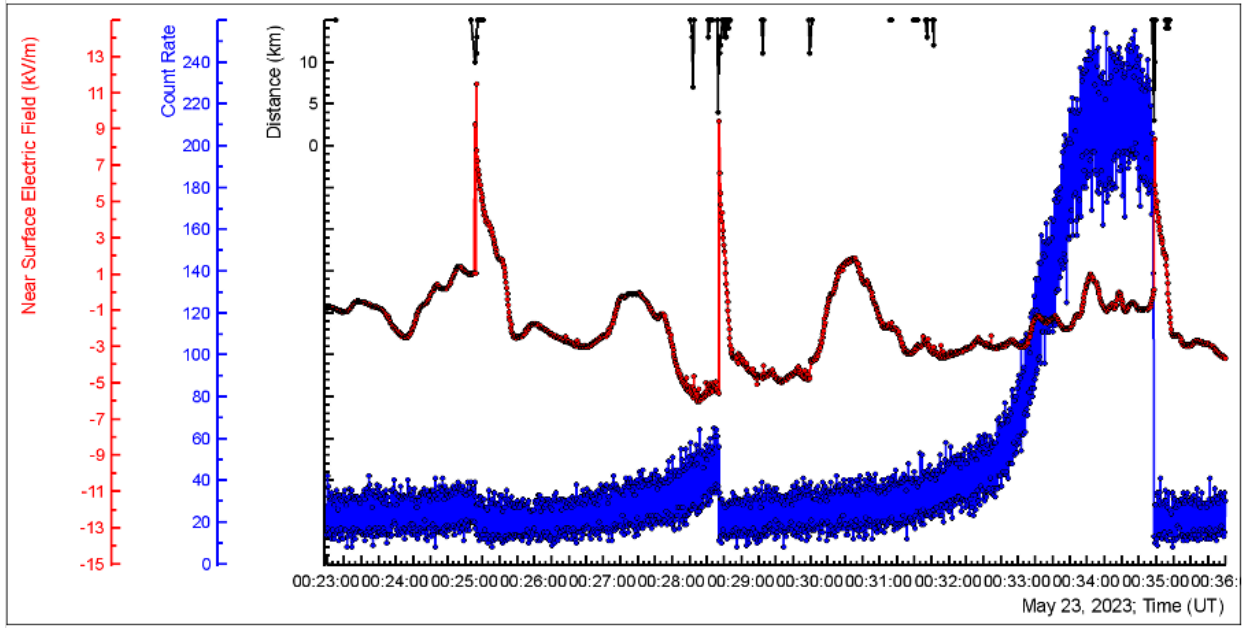

**Figure 4. 50 ms time series of 1 cm thick 1 area plastic scintillator, blue; disturbances of near-surface electric field measured by BOLTEK's EFM 100 electric mill, red; distances to the lightning flash, black.**

We estimate the particle flux above 0.3 MeV at maximum TGE flux to be 3,2 mln/min/m$^2$ and the fluence ≈ 500 particles/cm$^2$ (50 particles/cm$^2$ > 1MeV). On Lomnicky Stit in Slovakia, the particle flux was ≈ ten times larger [35]. In large-scale electric fields, free electrons accelerate and multiply, forming RREAs. These RREAs can reach space detectors and be registered as TGFs, and they can reach the ground and be registered as TGEs. Essentially, TGEs and TGFs are equivalent phenomena and can be considered two sides of the same coin. TGEs have been extensively researched, with hundreds of papers published and several datasets highlighting different aspects of their physics available to the public. Energy spectra have been recovered, intensities calculated, and their relation to lightning flashes and thundercloud's charged structure has been established. The lightning flashes stop particle fluxes, not producing additional seed electrons for RREAs. Figure 5 demonstrates that 5 TGEs occurred on a stormy day accompanied by multiple nearby lightning flashes. You can see that particle fluxes enhance between lightning flashes that abruptly terminate TGEs!

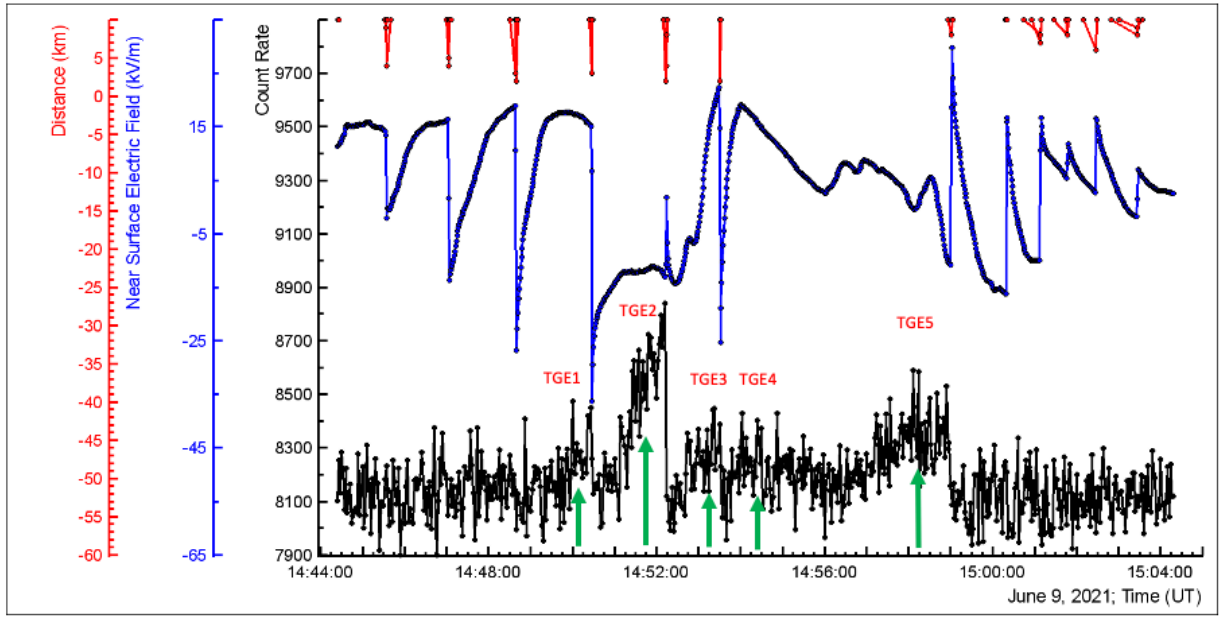

**Figure 5. 1-minute time series of count rates of ASNT particle detector, black; disturbances of the NSEF, blue; distances to lightning flashes, red.**

## 5. Cores of the Extensive air showers (EASs) as the origin of particle bursts

The physics of particle burst is associated with both the EAS phenomenon and HEPA (operation of electron accelerators in thunderclouds). The TA experiment relates these bursts to lightning activity. However, recent observations put the start of TGF before the lightning flash, and numerous minute-long TGEs also develop at the depressed lightning activity. In the previous section, we demonstrate the equivalence of TGE and TGF, which share the origination mechanism. There is also another, not well known to the community, mechanism that originates short particle bursts on the Earth's surface. Most of the EAS highest energy particles are located in the EAS core around the trajectory of the primary particle. The number of particles around the EAS axes (in a circle with radii < 1 m) can reach $10^6$ or more for primary protons with PeV energies. These particles come within a few tens of nanoseconds. A single plastic scintillator from the surface array with a usual dead time of 1 μs will generate only one large pulse in response to multiple particles from the EAS core and miss particle bursts. However, due to interactions of relativistic hadrons and gamma rays with soil and detector matter, it will be possible to register particle bursts with "slow" detectors. For detecting bursts of thermal neutrons, the HAWC experiment uses a scintillator (7.62 × 7.62 cm LaBr3 [10]). In Yangbajing, LHAASO, and URAN experiments [36-38], thermal neutrons were measured with electron-neutron (EN) detectors [39]. On the Aragats and Tien Shan mountains, neutron detectors were used for burst detection [40,41]. Below, we describe the burst detection with Aragats neutron monitor (ArNM).

The ArNM consists of 18 gas-filled cylindrical proportional counters of CHM-15 type (length 200cm, diameter 15cm) enriched with boron trifluoride ($^{10}BF_3$). The proportional counters are surrounded by 5 cm thick lead and 2 cm thick polyethylene covers. The cross-section of the lead above each section has a surface area of 6m$^2$, and the total surface area of the three sections is 18 m$^2$. In the inset to Fig. 6, we show one section of ArNM. The high-energy hadrons and gamma rays from EASs produce multiple neutrons in the lead. Then, the neutrons were thermalized in the polyethylene, entered the sensitive volume of the proportional counter, and yielded Li$^7$ and α particles via interactions with boron trifluoride [42]. The α particle accelerates in the high electrical field inside the proportional counter and produces a pulse registered by the DAQ electronics. Suppose only the incident hadrons must be measured (a one-to-one relationship between count rate and hadron flux). In that case, the dead time must equal the secondary neutron collection time (≈1250 µs) to avoid double-counting. If all pulses need to be counted, the dead time of the NM should be maintained very small. Yuri Stenkin et al., for the first time, described the detection of neutron bursts in the NM related to the occasional hitting of the detector by a core of a high-energy EAS [43]. Hadrons and gamma rays from the EAS core generate numerous thermal neutrons, increasing the NM count rate (neutron multiplicity). This option of EAS core detection by NM was almost not recognized in the past because the usually used long dead time does not permit counting the neutron multiplicity. By establishing a 3000 times shorter dead time of 0.4 µs, we detect EASs hitting ArNM, several of which provide bursts with a neutron multiplicity exceeding 6500, see Fig 6. The primary particle energies corresponding to this event should be very high (>10 PeV).

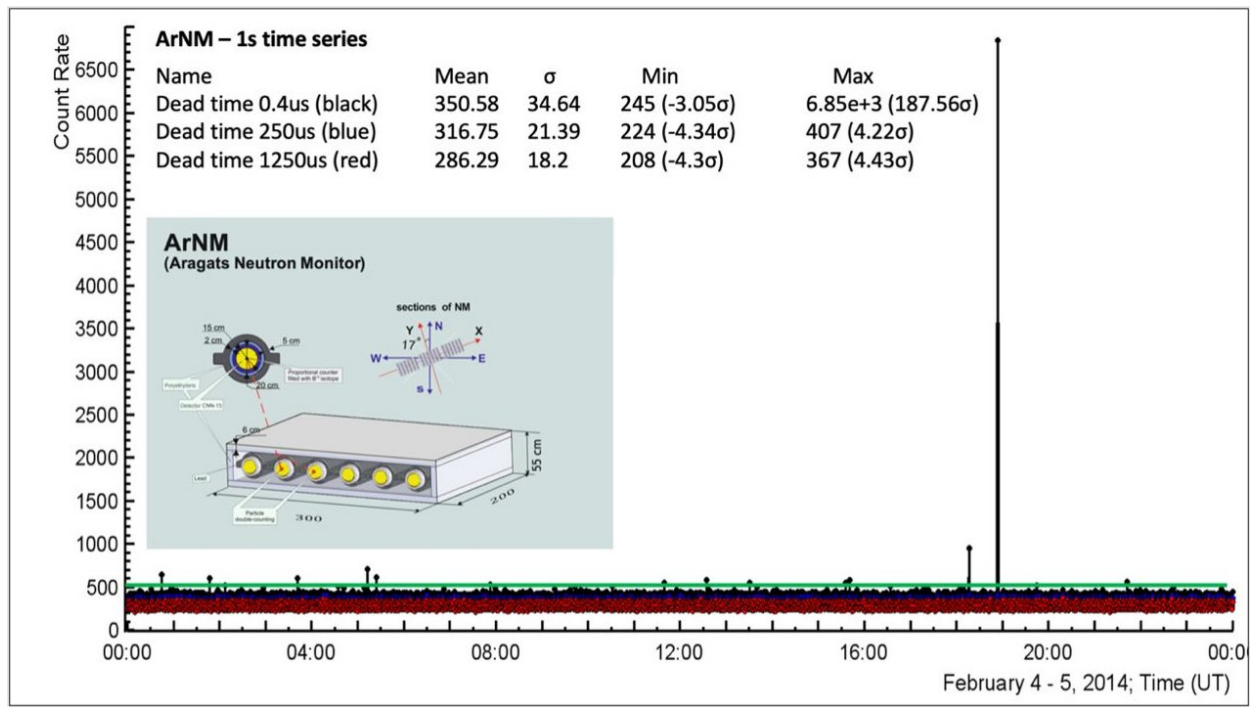

**Figure 6. 1s time series of ArNM for three different dead times; in the inset is shown the layout of one section of the 18NM64 neutron monitor. The green line outlines 13 bursts with multiplicity above 500, corresponding to a significance of 5σ relative to the mean value of the ArNM count rate.**

The signal of the ArNM was relayed via a high-frequency digitizing oscilloscope to see the neutron burst in more detail. Bursts were observed as sequences of microsecond pulses temporally isolated from other pulses on a time scale of at least 100 microseconds. Exhausting information on EAS core hitting ArNM can be found in the dataset of 50 high-multiplicity events published in the Mendeley repository[44]. From 50 selected bursts registered by ArNM, we obtain neutron burst durations of 2.6±0.6 ms. Approximately 100,000 times larger than the time needed for EAS core particles to cross the detector.

6. Discussion and conclusions

High altitude experiments HAWK, LHAASO, and ARGO-YBJ are looking for Pevatrons, detecting gamma rays from the point sources with energies of $10^{15}$ ev in excess. Energy estimation problems are crucial because the number of PeV particles detected can be no more than 1-2 for each source. Thus, enhancing the number of electrons in an atmospheric electric field can lead to energy overestimation and biases in the declaration of a Pevatron. In a simulation study [45], the results of which are shown in Table 1, the energy of a primary gamma-ray used in the simulation and calculated using the "measured" number of electrons (shower size) after EAS crosses the atmospheric electric field (as an energy estimator, a simple linear function was used that connected the shower size and the gamma-ray energy). As can be seen from the Table, the calculated energy of the primary gamma rays differs significantly from the "true" values. Thus, for the low primary energies (1-10 TeV), the bias in primary gamma-ray energy estimation can be tens-fold higher, for 100 TeV – 6.5 times, and for the higher energies (1 PeV) ≈2.5 times.

Table 1. "True" and estimated energies of primary gamma rays after passing through the electric field of 2.1 kV/cm strength.

| $E_o$ (GeV) | $E_{est}$ (GeV) |
|---|---|
| $10^3$ | $2.23 \times 10^4$ |
| $10^4$ | $1.34 \times 10^5$ |
| $10^5$ | $6.50 \times 10^5$ |
| $10^6$ | $2.42 \times 10^6$ |

All experiments (HAWK, LHAASO, ARGO-YBJ, TA, AUGER) report a significant influence of the atmospheric electric fields on the EAS trigger rate and CR intensity coinciding with the movement of the thundercloud. Atmospheric electric fields have less influence on experiments located on sea level (TA, AUGER) compared with high-mountain experiments. The atmospheric electric fields above surface arrays located at sea level are possibly too high. Therefore, the modulation effects are smaller than at mountain altitudes where the electric field can be prolonged to 50-150 m above the ground. The SEVAN detectors located at sea level in Hamburg and Berlin also didn't register TGEs. The exception is the TGEs (gamma glows) observed on the eastern coast of Japan in Winter [46]. However, due to specific climatic conditions, the clouds and electric fields are very low above the earth's surface.

TA and Pierre Auger experiments reported large-scale (≈200 km$^2$) short particle bursts originated 1-2 km above the Earth's surface; they call them DTGF in analogy to particle bursts detected by the orbiting gamma observatories. But only a handful of DTGFs have ever been detected on the ground, compared with many thousands in space. TGFs are short because of the experimental arrangement located the detector is 400-600 km away from the particle source on the fast-moving satellite. Few from millions of RREA particles born at altitudes ≈10 km in violent tropical storms reach orbiting detectors; sensors on balloons or aircraft flying just above the equatorial storm will register as many particles as TGEs do. The huge area covered by the particle bursts and the low location of their sources requires enormous amounts of gamma rays to originate very low above the ground. A "standard" upward TGF (used in simulations) produces about $10^{17}$ gamma rays of energy > 1 MeV in the primary beam [23]. A similar beam is used to simulate downward TGFs. However, the model of a highly bright point source of gamma rays 1-2 km above, if its footprint is 200 km, as at Auger observatory, will request a much brighter source to cover such an enormous area. Placing a source with fluence > $10^{20}$ gamma rays 1-2 km above ground is not the same as placing such a source in the open space. It will be another Oh-My-God particle observed by the Fly's Eye detector in 1991 (energy ≈50 J) or even more energetic.

The large footprint area possibly can be explained by the large sizes of the atmospheric electric field covering the detectors from above. TA and Auger surface arrays report on the influence of atmospheric electric fields on the count rates of particle detectors at huge areas following the movement of the thundercloud [13,15]. The atmospheric electric field, if its strength surpasses the critical value, can unleash large TGEs that can be detected by horizontally distributed detectors spanning tens of kilometers [47-50]. Thus, the particle bursts can be explained by the TGEs registered by remote detectors. They can coincide with atmospheric discharges or not; the driving force of enhanced particle fluxes is the prolonged large atmospheric electric field and not the atmospheric discharges of any type. All particle burst phenomena (TGF, DTGF, TGE) can be explained using the RREA model. The "lightning origin" of the bursts confuses the reader with too many uncertainties and contradictions [51]. Vast experimental data on TGEs and the time sequence between TGFs and lightning occurrence [29] confirms the universality of the RREA mechanism. Invoking, poorly understood lightning origin of the particle bursts violates the sufficient reason principle (principium rationis sufficientis cognoscendi). Each meaningful preposition can be considered reliable only if it is sufficiently grounded with arguments that can be regarded as valid. The "lightning" origin of particle bursts also contradicts Occam's razor principle (Entia nonsunt multiplicanda praetor necessitate), introducing too many complicated entities (different types of atmospheric discharges) instead of explaining the particle burst phenomena with more straightforward and well-known phenomena of RREAs and extensive air showers. The synergy between cosmic ray physics and atmospheric high-energy physics helps to explain the particle burst phenomena without invoking complicated and unproven modes of atmospheric discharges as the origin of particle bursts.

**ACKNOWLEDGMENTS** The author expresses gratitude to colleagues from the Cosmic Ray Division of Yerevan Physics Institute for their insightful discussions on TGE-TGF-Lightning problems. Additionally, thanks to F. Zagumenov for his work on sampling TGE events from oscilloscope datasets.